\documentclass[epj]{svjour}
\usepackage{graphics,amsmath,amssymb,multirow,color,pstricks,pst-plot}
\usepackage{xspace,graphicx,comment,cite}

\newcommand{\QQbar}{\ensuremath{Q\overline{Q}}\xspace}
\newcommand{\pt}{\ensuremath{p_{\rm T}}\xspace}
\newcommand{\dd}{\ensuremath{{\rm d}}\xspace}
\newcommand{\jpsi}{\ensuremath{{\rm J}/\psi}\xspace}
\newcommand{\psip}{\ensuremath{\psi{\rm (2S)}}\xspace}
\newcommand{\upsOne}{\ensuremath{\Upsilon{\rm (1S)}}\xspace}
\newcommand{\upsTwo}{\ensuremath{\Upsilon{\rm (2S)}}\xspace}
\newcommand{\upsThree}{\ensuremath{\Upsilon{\rm (3S)}}\xspace}

\begin{document}

\title{The fate of quarkonia in heavy-ion collisions at LHC energies:\\
a unified description of the sequential suppression patterns}

\titlerunning{The fate of quarkonia in heavy-ion collisions at LHC energies}

\author{Pietro Faccioli\inst{1} 
\and Carlos Louren\c{c}o\inst{2}
}
\institute{LIP and IST, Lisbon, Portugal
\and 
CERN, Geneva, Switzerland}

\date{Received: 27 July 2018 / Accepted: 3 September 2018}
\abstract{
Measurements made at the LHC have shown that 
the production of the \jpsi, \psip, \upsOne and \upsTwo quarkonia 
is suppressed in Pb-Pb collisions, 
with respect to the extrapolation of the pp production yields.
The \psip and \upsTwo states are more strongly suppressed than the ground states
and the level of the suppression changes with the centrality of the collision.
We show that the measured patterns can be reproduced by a simple model,
where all quarkonia are treated in a unified way, 
starting from the recent realisation that, in pp collisions,
the probability of quarkonium formation
has a universal dependence on the binding-energy of the bound state.
The hot-medium suppression effect is parametrized by a penalty factor in the binding energy,
identical for all (S- and \mbox{P-wave}) charmonium and bottomonium states, 
including those that indirectly contribute to the measured results through feed-down decays.
This single parameter, computed through a global fit of all available suppression patterns,
fully determines the hierarchy of nuclear effects, for all states and centrality bins.
The resulting faithful description of the data 
provides convincing evidence in favour of the conjecture of sequential quarkonium suppression induced by QGP formation.
\PACS{
      {12.38.Aw}{General properties of QCD}   \and
      {12.38.Qk}{Experimental tests of QCD}   \and
      {12.38.Mh}{Quark-gluon plasma}
     } 
}

\maketitle

\section{Introduction}

The theory of strong interactions, quantum chromodynamics (QCD), 
predicts the existence of a deconfined system of quarks and gluons (quark gluon plasma, QGP),
formed when the QCD medium reaches a sufficiently-high temperature.
To produce and study this extremely hot state of matter, 
experiments collide heavy nuclei at the highest possible energies 
and look for significant modifications in the rates and distributions of the produced particles, 
with respect to baseline properties measured in proton-proton collisions. 
One of the proposed signatures of QGP formation is that 
quarkonium bound states should be produced less and less frequently, 
as the binding potential between the constituent heavy quark and antiquark is screened 
by the colour-charge distribution of the surrounding quarks and gluons~\cite{bib:MS}. 
The distinctive feature of this effect is its ``sequentiality'': 
the suppression of the production of different quarkonium states should happen progressively, 
as the temperature of the medium increases,
following a hierarchy in binding energy~\cite{bib:DPS, bib:KKS}.

Obtaining convincing evidence of this sequential mechanism is a challenge to experiments. 
Only a small number of the many quarkonium states, 
in their variety of flavour compositions, masses, binding energies, sizes, angular momenta and spins, 
have been observed in nucleus-nucleus collisions.
For several of them, even the baseline proton-proton production rates and kinematics are poorly known. 
Furthermore, even for the most copiously produced and best-known states, \jpsi and \upsOne, 
it is not well known how many of them are directly produced. 
The (large) fraction resulting from decays of heavier states must be precisely evaluated and accounted for, 
since it is the ``mother'' particle, with its specific properties, that undergoes the interaction with the QCD medium.  
Moreover, measurements of different states are not always directly comparable, 
because of differences in the kinematic phase space windows covered by the detectors 
or because of inconsistent choices in the binning of the published results.
On the theory side, the interpretation of the data in terms of the ``signal'' sequential suppression effect 
is obfuscated by a variety of (hypothetical) ``background'' medium effects, 
such as quarkonium formation from initially-uncorrelated quarks and antiquarks, 
break-up interactions with other particles, energy loss in the nuclear matter, 
modifications of parton distributions inside the nuclei, etc.

We present a data-driven model that considers quarkonium suppression
from proton-proton to nucleus-nucleus collisions through a minimal modification
of the ``universal'' (state-independent) patterns recently observed in pp data~\cite{bib:EPJCscaling}. 
It is based on a simple and single empirical hypothesis:
the mechanism of nuclear modification depends \emph{only} on the quarkonium binding energy, 
with no distinction between the charmonium and bottomonium families, 
nor between states of different masses and spins. 
The model is used to fit the nuclear modification factors, $R_{\rm AA}$,
measured by CMS and \mbox{ATLAS} at $\sqrt{s} = 5.02$\,TeV,
in bins of collision centrality defined using the number of participant nucleons, $N_\mathrm{part}$.
The result of this global fit, using the most detailed and precise measurements currently available,
is that a simple hierarchy in binding energy can explain the observed quarkonium suppression patterns.
In other words, the presently available data provide a clear signature of the sequential suppression conjecture,
according to which 
the more strongly-bound states are progressively suppressed as the temperature of the medium exceeds certain thresholds.

\section{Quarkonium suppression patterns}
\label{sec:model}

\begin{figure}[b]
\centering
\includegraphics[width=0.72\linewidth]{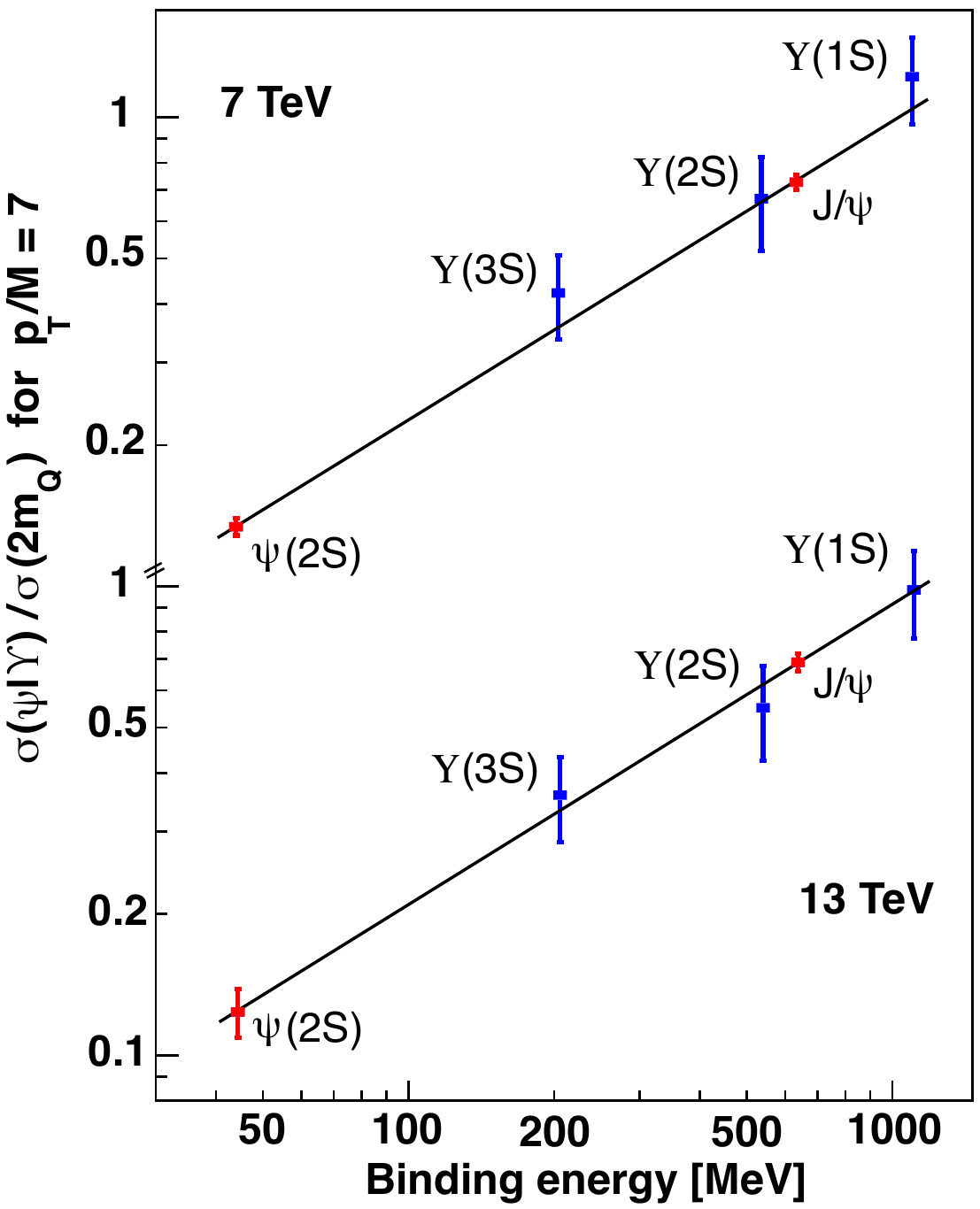}
\caption{Direct production cross sections of quarkonia in pp collisions 
(normalized to the extrapolated cross section of a state of mass $2m_Q$), 
shown as a function of $E_{\mathrm{b}}$ at 7 and 13\,TeV.}
\label{fig:ppEbindingScaling}
\end{figure}

At the current level of experimental precision, 
the \pt-differential charmonium and bottomonium production cross sections
measured in 7 and 13\,TeV pp collisions 
at mid-rapidity~\cite{bib:ATLASYnS,bib:ATLASpsi2S,bib:ATLASchic,bib:CMSjpsi,bib:CMSYnS,bib:BPH15005}
are well reproduced by a simple parametrization reflecting a 
universal (state-independent) energy-momentum scaling~\cite{bib:EPJCscaling}. 
In this description, the \emph{shape} of the mass-rescaled transverse momentum ($p_{\rm T}/M$)
distribution is independent of the quarkonium state, 
while its \emph{normalization} 
(at any chosen $p_{\rm T}/M$ value) 
shows a clear correlation with the binding energy, 
calculated as the difference between the open-flavour threshold and the quarkonium mass,
$E_{\mathrm{b}} = 2M(D^0) - M(\psi{\rm (nS)})$ or $2M(B^0) - M(\Upsilon{\rm (nS)})$.
The observed correlation, shown in Fig.~\ref{fig:ppEbindingScaling},
is seemingly identical for the charmonium and bottomonium families,
and for the two collision energies. 

The linear correlation seen in the log-log representation of Fig.~\ref{fig:ppEbindingScaling}
suggests that we can faithfully parametrize the $E_{\mathrm{b}}$ dependence of the 
\mbox{S-wave} direct-production cross sections using a power-law function:
\begin{equation}
f_{\mathrm{pp}}^{\psi/\Upsilon}(E_{\mathrm{b}}) \equiv
\left( \frac{\sigma^{\mathrm{dir}}(\psi/\Upsilon)}{\sigma(2m_Q)} \right)_{\mathrm{pp}}
 = \left( \frac{E_{\mathrm{b}}}{E_0}\right)^{\delta} \quad .
\label{eq:ppEbindingScaling}
\end{equation}

Here, $\sigma(2m_Q)$ is the extrapolation (at fixed $p_{\rm T}/M$) of the cross section 
to twice the relevant heavy quark mass, 
computed from the mass of the lightest quarkonium state:
$2m_c = M(\eta_c)$ and $2m_b = M(\eta_b)$~\cite{bib:PDG}.
One single exponent parameter $\delta$ is used for both quarkonium families,
so as to minimize the number of free parameters in the model,
especially in view of the uncertainties of some experimental measurements.
Independent fits at the two collision energies give the values
$\delta = 0.63 \pm 0.02$ at 7\,TeV and $0.63 \pm 0.04$ at 13\,TeV~\cite{bib:EPJCscaling}.
The equation defines a universal ``bound-state transition function'', 
$f_{\mathrm{pp}}^{\psi/\Upsilon}(E_{\mathrm{b}})$,
proportional to the probability that the \QQbar pre-resonance
evolves to a given $\psi/\Upsilon$ state. 
The transition process involves long-distance interactions between the quark and the antiquark,
for which no theory calculations exist.

Given the current lack of \mbox{P-wave} cross section data,
we assume that the direct production of $\chi_c$ and $\chi_b$ is described by an analogous bound-state transition function,
with identical dependence on the binding energy (same $\delta$ value as for the S-wave states),
but an independent $E_0$ value (reflecting the different angular momentum and wave-function shape).
Complementing this long-distance scaling with the short-distance production ratio
$\sigma(2m_b) / \sigma(2m_c) = (m_b / m_c)^{-6.63 \pm 0.08}$~\cite{bib:EPJCscaling}, 
we obtain a complete parametrization of the direct production cross sections
for all states of the charmonium and bottomonium families. 
Together with the relevant feed-down branching fractions~\cite{bib:PDG}, 
this provides a full picture of inclusive quarkonium production in pp collisions, 
including the detailed contributions of the feed-down decays from heavier to lighter states, 
as reported in Ref.~\cite{bib:EPJCscaling}.
This data-driven model is used in the present study as a baseline for the interpretation of the Pb-Pb data.

\begin{figure}[t]
\centering
\includegraphics[width=0.75\linewidth]{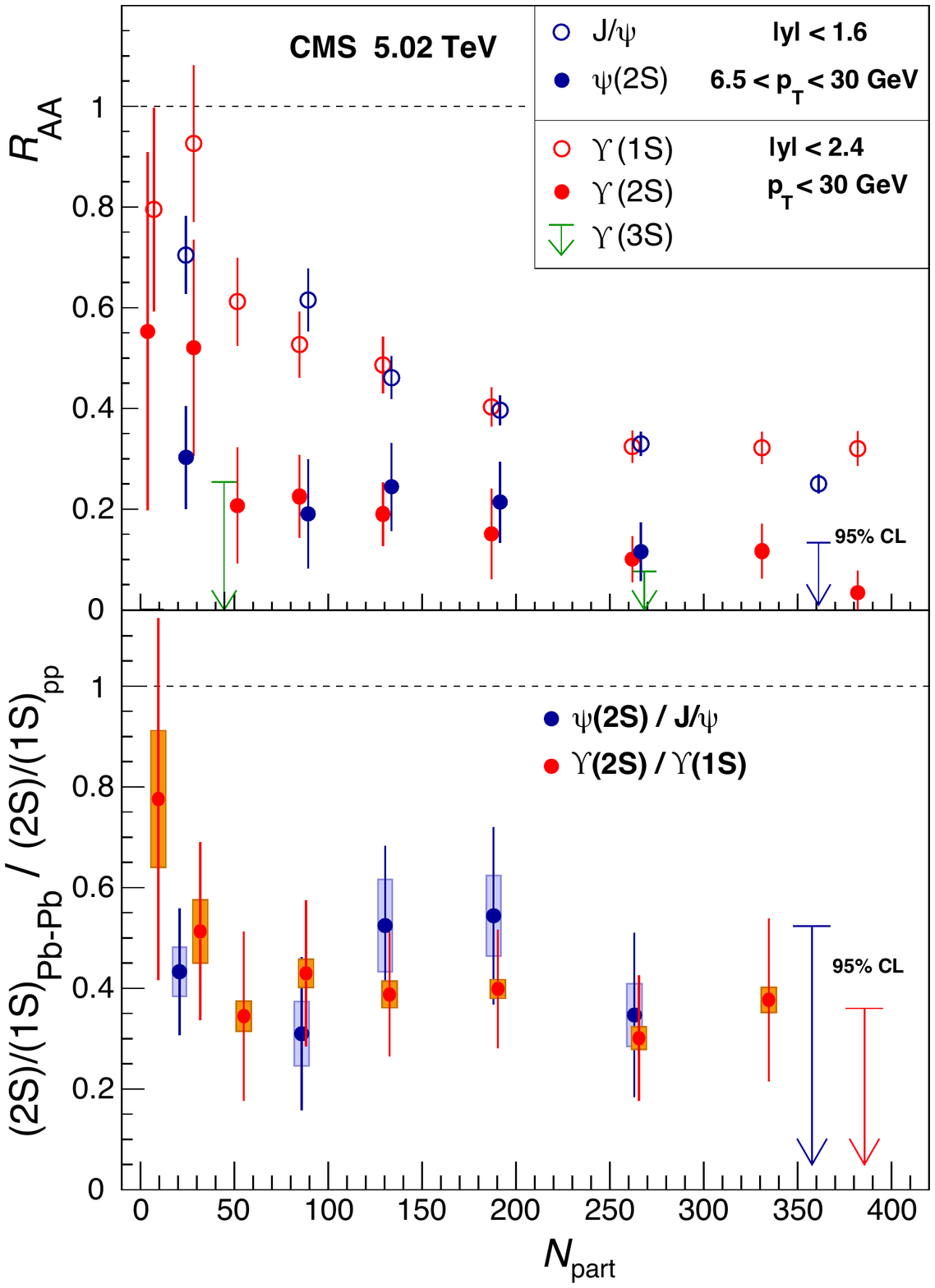}
\caption{Top: Nuclear modification factor as a function of centrality for the 
(inclusive) \jpsi, \psip, \upsOne, \upsTwo and \upsThree quarkonia, as measured by CMS
comparing pp and Pb-Pb data at 5.02\,TeV~\cite{bib:CMS_RAA_psi,bib:CMS_RAA_upsilon}.
Bottom: Corresponding double ratio of the 2S and 1S nuclear modification factors.}
\label{fig:RAA_DRs_Data}
\end{figure}

Our hypothesis on how the pp baseline is modified in Pb-Pb collisions is guided by the  
experimental observation that the \psip and \jpsi exhibit very different suppression patterns 
in Pb-Pb collisions as a function of collision centrality~\cite{bib:CMS_RAA_psi},
as shown in Fig.~\ref{fig:RAA_DRs_Data}-top.
The $R_{AA}$ of the \psip shows a significant departure from unity
already in the most peripheral bin probed by the experiments, 
corresponding to an average number of colliding nucleons of $N_\mathrm{part} = 22$,
and then seems to be almost independent of $N_\mathrm{part}$ up to the most central Pb-Pb collisions.
Instead, the $R_{AA}$ of the \jpsi shows a more gradual decrease from peripheral to central collisions,
being relatively close to unity in the most peripheral bins.
We can also see in Fig.~\ref{fig:RAA_DRs_Data}-top that 
the \upsOne and \upsTwo suppression patterns~\cite{bib:CMS_RAA_upsilon} are very similar 
to those of the \jpsi and \psip, respectively.

The different suppression patterns of the 2S and 1S states can also be appreciated through the
double suppression ratios ($R_{AA}(\mathrm{2S}) / R_{AA}(\mathrm{1S})$) measured by CMS,
as shown in Fig.~\ref{fig:RAA_DRs_Data}-bottom for the two quarkonium families.
The charmonium double ratio is significantly smaller than unity already in the most peripheral collisions,
confirming that the \psip is strongly suppressed even in the most ``pp-like" nuclear collisions.
The \jpsi and \psip suppression patterns reported by ATLAS~\cite{bib:ATLAS_RAA} show similar
features.

The apparent fragility of the \psip can be attributed to its binding energy, 44\,MeV, 
very small with respect to both its mass and the open charm mass threshold, $2M(D^0)$.
A fluctuation of around 1\% in the invariant mass of the pre-resonance \QQbar
or in the threshold energy above which open charm production becomes possible
is sufficient to inhibit the formation of this weakly-bound quarkonium state.
This concept can be formalized through a minimal modification of the pp production baseline, 
in which the short-distance partonic production of the \QQbar state is assumed to remain unchanged, 
while the long-distance bound-state transition function (Eq.~\ref{eq:ppEbindingScaling}) becomes
\begin{equation}
f_{\mathrm{PbPb}}^{\psi/\Upsilon}(E_{\mathrm{b}} , \epsilon)  \equiv
\left( \frac{\sigma^{\mathrm{dir}}(\psi/\Upsilon)} {\sigma(2m_Q)} \right)_{\mathrm{PbPb}} = 
\left( \frac{E_{\mathrm{b}}-\epsilon}{E_0} \right)^{\delta}
\label{eq:AAEbindingScaling}
\end{equation}
 for $E_{\mathrm{b}} - \epsilon > 0$ and vanishes for $E_{\mathrm{b}} - \epsilon < 0$. 
Here, $\epsilon$ represents a shift in the difference between the di-meson threshold energy and the \QQbar mass. 
The magnitude of $\epsilon$ measures the strength of the observable nuclear suppression effects: 
as $\epsilon$ increases it becomes progressively less probable to \emph{form} the bound state
and once $\epsilon$ exceeds $E_{\mathrm{b}}$ the \QQbar pair never binds into a quarkonium state.

This empirical parametrization implicitly reflects different possible physics effects.
For example, multiple scattering effects may increase on average the relative momentum 
and invariant mass of the unbound quark and antiquark~\cite{bib:JWQiu_suppression},
pushing such pairs towards or beyond the di-meson threshold. 
Alternatively, or simultaneously, 
a screening of the attractive interaction between the quark and the antiquark
may disfavour the formation of a bound state,
tending to separate the two objects and ultimately leading to two independent hadronizations. 
Both examples can be described in this model, assuming $\epsilon > 0$.

We indicate with $\langle \epsilon \rangle$ and $\sigma_\epsilon$ the average and width 
of the $\epsilon$ distribution 
characterizing a given experimental condition, 
mainly defined by the collision energy and the centrality-distribution of the events.
Correspondingly, we define the event-averaged bound-state transition function   
\begin{equation}
F_{\mathrm{PbPb}}^{\psi/\Upsilon}( E_{\mathrm{b}}, \langle \epsilon \rangle, \sigma_\epsilon) = \frac{\int_{0}^{ E_{\mathrm{b}}} [(E_{\mathrm{b}}-\epsilon)/E_0]^{\delta} \; G(\epsilon; \langle \epsilon \rangle, \sigma_\epsilon) \; \dd\epsilon}{\int_{0}^{E_{\mathrm{b}}} G(\epsilon; \langle \epsilon \rangle, \sigma_\epsilon) \; \dd\epsilon} \quad,
\label{eq:transfunctionPbPb_avg}
\end{equation}
where $\epsilon$ is distributed following a function $G$, assumed, for simplicity, to be Gaussian.

The resulting nuclear suppression ratio for \emph{direct} quarkonium production 
is calculated in this model as the ratio between the long-distance bound-state transition functions 
of the Pb-Pb and pp cases:
\begin{equation}
R_{AA}^{\mathrm{dir}}( E_{\mathrm{b}}, \langle \epsilon \rangle, \sigma_\epsilon) = F_{\mathrm{PbPb}}^{\psi/\Upsilon}( E_{\mathrm{b}}, \langle \epsilon \rangle, \sigma_\epsilon) \; / \; f_{\mathrm{pp}}^{\psi/\Upsilon}(E_{\mathrm{b}}) \quad .
\label{eq:RAA_dir}
\end{equation}

In principle, the energy-shift effect, and therefore $\langle \epsilon \rangle$ and $\sigma_\epsilon$, 
may depend on the identity of the quarkonium state.
However, in line with the seemingly universal properties of quarkonium production in pp collisions, 
we will work under the hypothesis that also the suppression can be parametrized with a ``universal'' $\epsilon$ distribution, 
identical for all quarkonia. 
Throughout the following discussion this will remain our central hypothesis, 
which we want to test using the \jpsi, \psip, \upsOne and \upsTwo measurements.

\begin{figure}[t]
\centering
\includegraphics[width=0.85\linewidth]{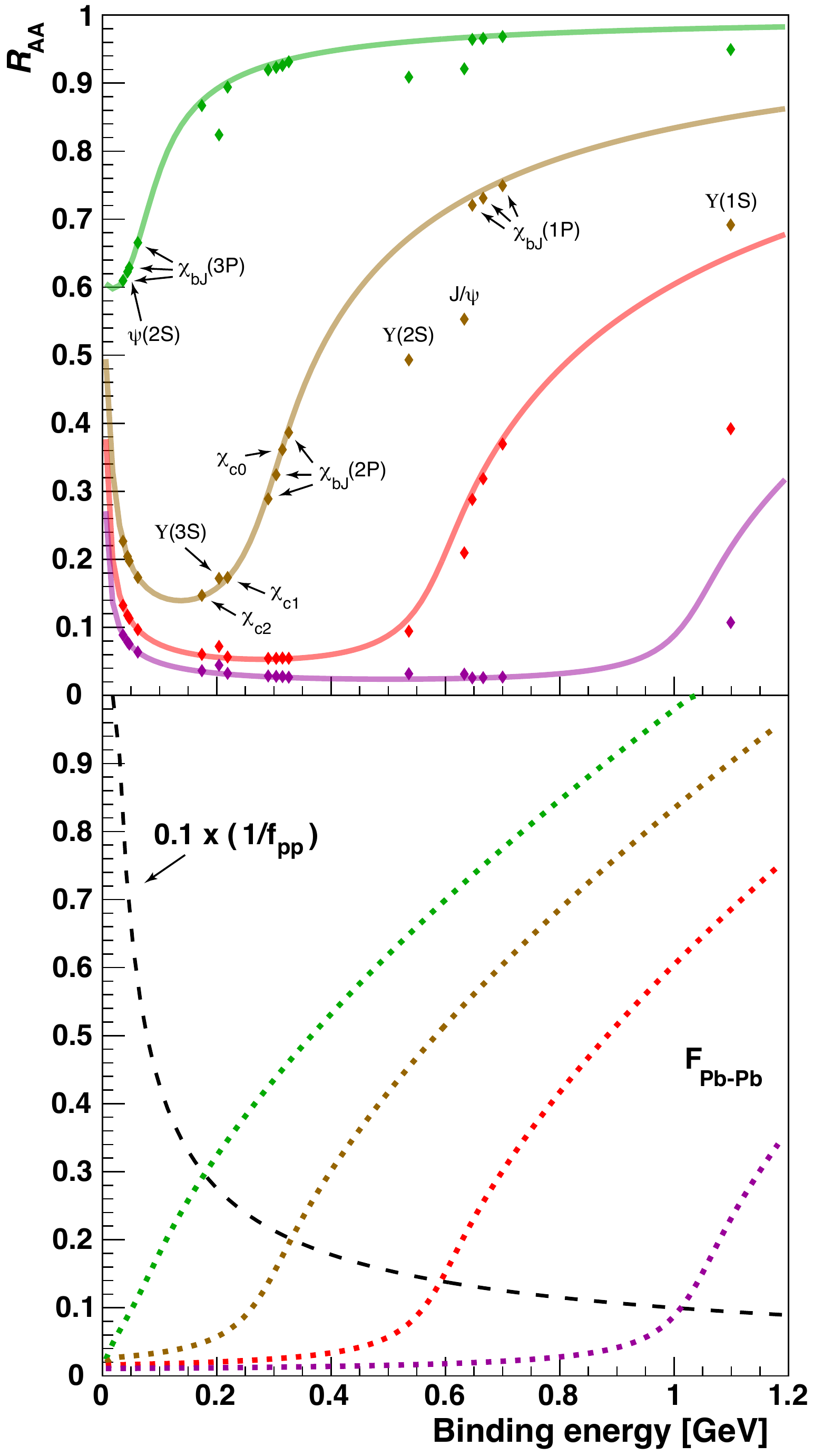}
\caption{Top: Calculated nuclear suppression factor as a function of the quarkonium $E_\mathrm{b}$
for direct (curves) and inclusive (markers) production of S- and \mbox{P-wave} states. 
Results are shown for $\sigma_\epsilon = 30$\,MeV and for
$\langle \epsilon \rangle = 20$, 250, 550 and 1000\,MeV,
respectively shown in green, brown, red and violet. 
Bottom: Individual contributions (Eq.~\ref{eq:RAA_dir}) 
of the long-distance bound-state transition functions in pp ($1/f_{\mathrm{pp}}^{\psi/\Upsilon}$, 
scaled by 0.1) and Pb-Pb ($F_{\mathrm{PbPb}}^{\psi/\Upsilon}$, 
for the same $\sigma_\epsilon$ and $\langle \epsilon \rangle$ values).}
\label{fig:RAA_computation}
\end{figure}

While $R_{AA}^{\mathrm{dir}}$ is defined continuously for any value of $E_\mathrm{b}$ 
(including values not corresponding to physical bound states), 
the nuclear suppression ratio for inclusive quarkonium production 
depends on the feed-down contributions specific to each observable state, 
therefore becoming a discreet set of points. 
We model the observable suppression for the quarkonium state $\psi_k$ as
\begin{equation}
\label{eq:RAA_prompt}
\begin{split}
& R_{AA}^{\mathrm{inc}}(\psi_k, \langle \epsilon \rangle, \sigma_\epsilon) = \\
& \frac{ 
\sum_{j} R_{AA}^{\mathrm{dir}}[E_{\mathrm{b}}(\psi_j),\langle \epsilon \rangle, \sigma_\epsilon] \;  
\sigma_{\mathrm{pp}}^{\mathrm{dir}}(\psi_j) \; {\mathcal B}(\psi_j \to \psi_k)  }
{\sum_{j}  \sigma_{\mathrm{pp}}^{\mathrm{dir}}(\psi_j) \; {\mathcal B}(\psi_j \to \psi_k) } \, ,
\end{split}
\end{equation}
where, according to the hypothesis that the observed suppression is driven by a state-independent energy-shift effect, 
$\langle \epsilon \rangle$ and $\sigma_\epsilon$ do not depend on $j$ and $k$. 
Naturally, ${\mathcal B}(\psi_j \to \psi_k) = 0$ if $m(\psi_k) > m(\psi_j)$ and ${\mathcal B}(\psi_j \to \psi_j) = 1$.

For $\sigma_{\mathrm{pp}}^{\mathrm{dir}}(\psi_j)$ 
we use the full set of direct production cross sections determined, as mentioned above, 
in the global parametrization of mid-rapidity 7\,TeV data~\cite{bib:EPJCscaling}. 
The choice of a specific energy for the pp reference does not affect $R_{AA}^{\mathrm{inc}}$ 
as long as the cross section \emph{ratios} (as effectively appearing in Eq.~\ref{eq:RAA_prompt}) 
do not depend on the pp collision energy, 
an hypothesis fully consistent with the scaling properties discussed in Ref.~\cite{bib:EPJCscaling}.

The $E_{\mathrm{b}}$ dependence of $R_{AA}^{\mathrm{dir}}$ 
is shown in Fig.~\ref{fig:RAA_computation}-top as four continuous curves,
corresponding to different $\langle \epsilon \rangle$ values, while the 
corresponding $R_{AA}^{\mathrm{inc}}$ discrete values, 
accounting for the state-specific feed-down contributions, are shown as sets of points,
placed at the binding energies of the physical quarkonium states, reported in Table~\ref{tab:EB}.
In this preliminary illustration we have not shown the effect of the uncertainties affecting $\delta$, 
the branching ratios, and the cross sections.
The $R_{AA}^{\mathrm{dir}}$ continuous curves show a strong suppression up to 
$E_{\mathrm{b}} \simeq \langle \epsilon \rangle$, 
followed by a power-law increase (determined by $\delta$).
This shape reflects the behavior of the Pb-Pb bound-state transition function,
modelled in Eqs.~\ref{eq:AAEbindingScaling} and~\ref{eq:transfunctionPbPb_avg}, 
and shown in Fig.~\ref{fig:RAA_computation}-bottom.

\begin{table}[t]
\centering
\caption{Binding energies of the quarkonia shown in Fig.~\ref{fig:RAA_computation}.}
\label{tab:EB}
\begin{tabular}{cccc}
\hline\noalign{\vglue0.85mm}
Quarkonium & $E_\mathrm{b}$\,[MeV] &
Quarkonium & $E_\mathrm{b}$\,[MeV] \\
\noalign{\vglue0.85mm}\hline\noalign{\vglue0.85mm}
$\chi_{b2}$(3P) &  36    & $\chi_{c0}$ &  315   \\
\psip     &  44                & $\chi_{b0}$(3P) &  326   \\
$\chi_{b1}$(3P) &  47   & $\Upsilon$(2S)       &  536   \\
$\chi_{b0}$(3P) &  62   & \jpsi      &  633   \\
$\chi_{c2}$ &  174        & $\chi_{b2}$(1P)  &  647   \\
$\Upsilon$(3S) &  204  & $\chi_{b1}$(1P) &  666   \\
$\chi_{c1}$ &  219        & $\chi_{b0}$(1P)  &  700   \\
$\chi_{b2}$(2P) &  290 & $\Upsilon$(1S) &  1099   \\
$\chi_{b1}$(3P) &  304 & & \\
\noalign{\vglue0.85mm}\hline
\end{tabular}
\end{table}

\begin{figure}[t]
\centering
\includegraphics[width=0.85\linewidth]{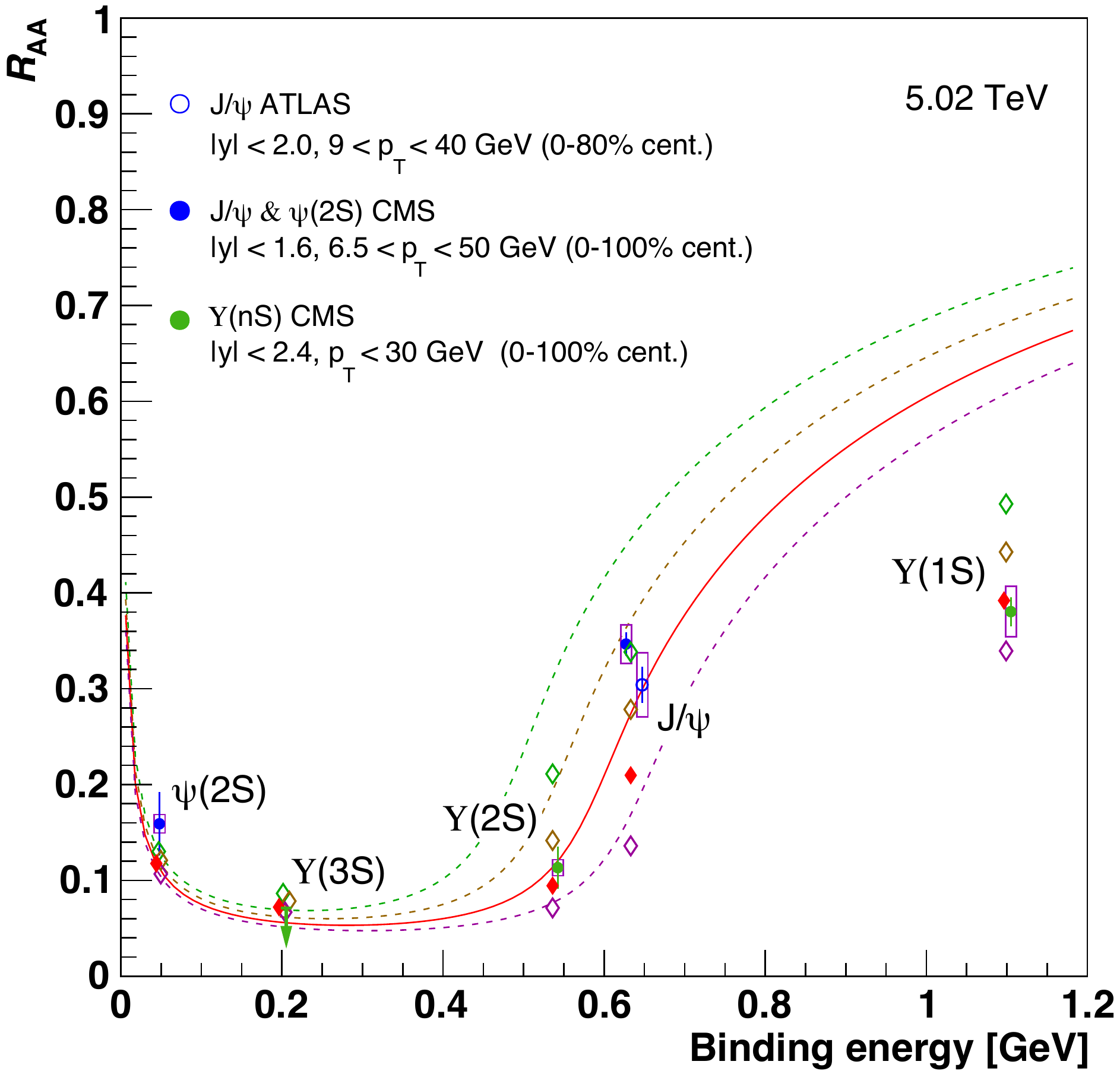}
\caption{Comparison between the measured 
values~\cite{bib:CMS_RAA_psi, bib:CMS_RAA_upsilon, bib:ATLAS_RAA}, 
integrated over the probed centrality range, 
and the results computed for $\langle \epsilon \rangle = 450$, 500, 550 and 600\,MeV,
respectively shown in green, brown, red and violet.}
\label{fig:RAA_examples}
\end{figure}

As illustrated in Fig.~\ref{fig:RAA_examples}, 
the model reproduces the observed hierarchy of centrality-integrated suppression values,
with a universal $\langle \epsilon \rangle$ around 550\,MeV.
The importance of accounting for the feed-down contributions is clearly seen in the
\jpsi and \upsOne cases, where $R_{AA}^{\mathrm{inc}}$ and $R_{AA}^{\mathrm{dir}}$ 
are particularly different.
The ATLAS \jpsi measurement does not include the most peripheral Pb-Pb collisions, 
which might explain the slightly smaller $R_{AA}$ value, relative to the CMS measurement.
While the $\Upsilon$(nS) data points are very well reproduced by the computation made with 
$\langle \epsilon \rangle = 550$\,MeV, the measured \jpsi and \psip $R_{AA}$ are higher than the
computed values. This might be an indication that charmonium production in Pb-Pb collisions 
at the LHC energies includes an extra contribution with respect to those at work in pp collisions, 
one option being the binding of uncorrelated quarks and antiquarks 
(produced in different nucleon-nucleon interactions), 
made possible by the very large number of charm quarks 
produced in these collisions~\cite{bib:Thews, bib:PBM}
and seemingly observed at low-\pt by ALICE~\cite{bib:ALICE}.
The level of this contribution should be significantly reduced in the \pt region probed by the 
CMS and ATLAS data, $\pt > 6.5$ and 9\,GeV, respectively, but it could well be that the residual 
contamination has a visible effect.

It is worth highlighting that the maximum-suppression plateau 
in the region $E_{\mathrm{b}} < \langle \epsilon \rangle$
becomes wider for larger values of $\langle \epsilon \rangle$, but $R_{AA}$ never vanishes. 
This effect is qualitatively determined by the non-zero width $\sigma_\epsilon$ of the energy-shift distribution. 
The use of a distribution for $\epsilon$ (instead of a fixed average value)
roughly simulates a realistic mixture of physical events where, for example, 
quarkonia produced in the scattering of nucleons in the nuclear halos (small $\epsilon$) 
can survive even in the most central collisions (large $\langle \epsilon \rangle$). 
The modelling of this ``tail'' effect and, therefore, of the shape of the plateau, 
reflects the shape of the $\epsilon$ distribution (here simply assumed to be a symmetric Gaussian). 
Future measurements of very small suppression factors close to the centre of the plateau 
(for example, the one of the \upsThree state, for which only upper limits exist so far) 
will probe different shape hypotheses. 
We also note that the increase of $R_{AA}$ as $E_{\mathrm{b}} \to 0$, 
determining the prediction that the \psip is \emph{less} suppressed than the \upsThree, 
is actually a ``pp effect'', 
caused by the presence of $f_{\mathrm{pp}} \propto E_{\mathrm{b}}^\delta$ 
in the denominator of $R_{AA}$ 
(Eqs.~\ref{eq:ppEbindingScaling} and ~\ref{eq:RAA_dir}), 
as illustrated in the bottom panel of Fig.~\ref{fig:RAA_computation}.

\section{Global fit of the $R_{AA}$ data}
\label{sec:fit}

Having introduced and motivated our model in the previous section, we will now move to a 
more quantitative analysis of the experimental data.
As mentioned before, 
the CMS and ATLAS Collaborations have reported quarkonium suppression measurements 
using pp and Pb-Pb collisions at 5.02\,TeV, in comparable experimental conditions,
for four different states, \jpsi, \psip, \upsOne and \upsTwo, 
complemented by upper limits for the 
\upsThree~\cite{bib:CMS_RAA_psi,bib:CMS_RAA_upsilon,bib:ATLAS_RAA}. 
We performed a global analysis of 37 $R_{AA}$ \jpsi, \psip, \upsOne and \upsTwo values,
measured in several $N_\mathrm{part}$ bins,
testing the hypothesis of a universal mechanism, 
the intensity of which is measured by an average shift $\langle \epsilon \rangle$ 
of the binding energy $E_\mathrm{b}$, 
common to all states and depending on $N_\mathrm{part}$. 

Preliminary fits for individual $N_\mathrm{part}$ bins show a definite correlation 
between $\langle \epsilon \rangle$ and the logarithm of $N_\mathrm{part}$, 
while no significant dependence of $\sigma_\epsilon$ on $N_\mathrm{part}$ is seen. 
Therefore, the global fit of all data is performed assuming a linear dependence of 
$\langle \epsilon \rangle$ on $\ln(N_\mathrm{part})$. 
The two coefficients of this dependence and the ($N_\mathrm{part}$-independent) $\sigma_\epsilon$ 
are the parameters of the fit.
While there is, a priori, no reason to assume that $\sigma_\epsilon$ is independent of $N_\mathrm{part}$,
the accuracy of the presently available measurements justifies approximating it by a constant, 
which provides a very good description of the data
and makes the global fit more robust than if we would have included more free parameters.
Other functional forms can be considered once more precise and detailed data will become available.

Two global uncertainties, common to all ATLAS or CMS data points, 
and four uncertainties correlating the CMS points of each quarkonium state 
are taken into account by introducing corresponding constrained (nuisance) parameters in the fit. 
Further nuisance parameters are $\delta = 0.63 \pm 0.04$, 
parametrizing the power-law dependence on $E_\mathrm{b}$, 
and several constrained factors used to model the uncertainties 
in the direct production cross sections and feed-down branching ratios 
entering Eq.~\ref{eq:RAA_prompt}. 
The latter uncertainties have a negligible impact in the results. 

\begin{figure}[!t]
\centering
\includegraphics[width=0.8\linewidth]{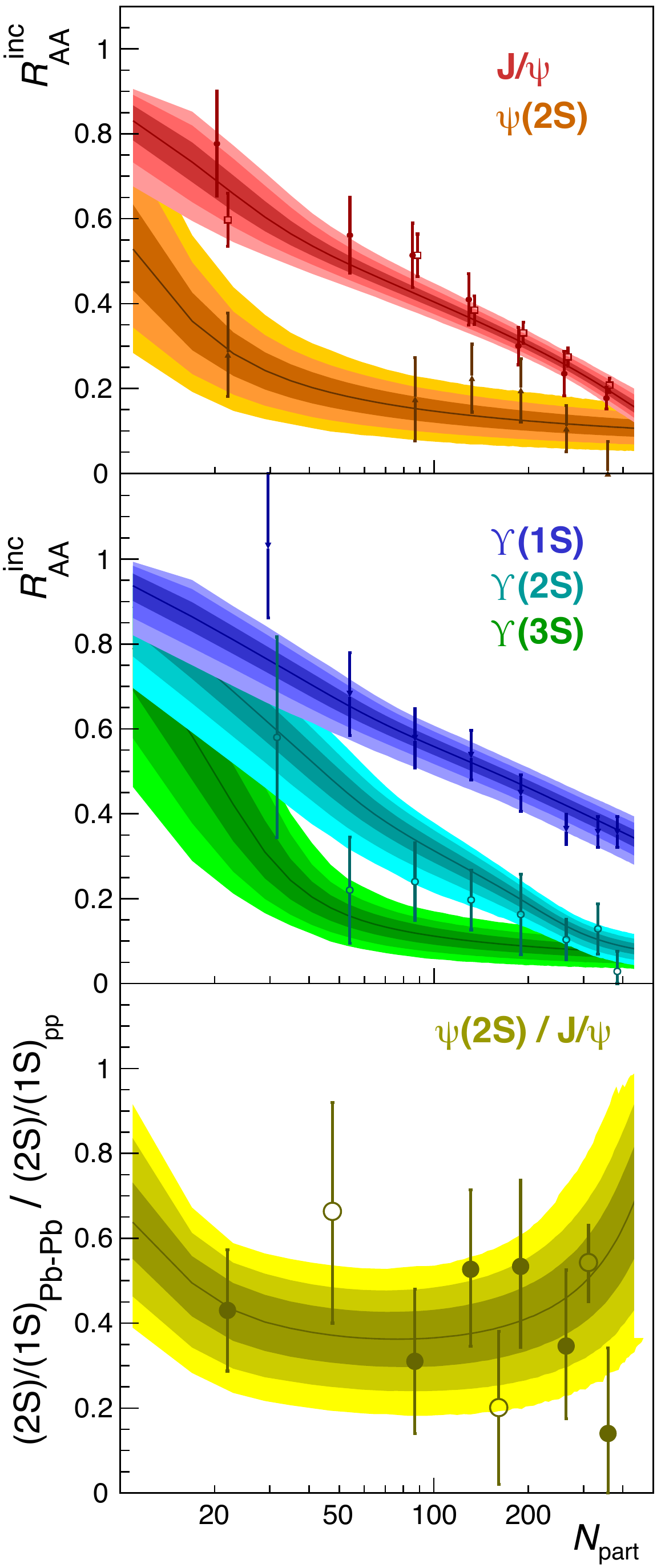}
\caption{$R_{AA}^{\mathrm{inc}}$ curves resulting from the global fit, 
with bands representing the 68, 95 and 99.7\% confidence level intervals, 
compared to the corresponding charmonium (top) and bottomonium (middle) 
measurements by CMS (closed markers) and ATLAS (open markers),
as a function of $N_\mathrm{part}$.
The corresponding \psip-to-\jpsi double ratios (not included as fit constraints)
are shown in the bottom panel.}
\label{fig:RAA_results_vs_Npart}
\end{figure}

Figure~\ref{fig:RAA_results_vs_Npart} shows how the fit results
(coloured bands) compare to the measurements, as a function of $N_\mathrm{part}$.
The fit has a high quality, with a total $\chi^2$ of 40 for 34 degrees of freedom,
and a 22\% probability that a higher $\chi^2$ value would be obtained 
if the data points were statistical fluctuations around perfectly modelled central values.
The measured (inclusive) \psip-to-\jpsi charmonium suppression double ratio, not included in the fit, 
is well reproduced in its seemingly inexistent $N_\mathrm{part}$ dependence. 
A strong deviation from the approximately flat behaviour should be observed 
through more precise and finely binned measurements in the low-$N_\mathrm{part}$ region.
\begin{figure}[t]
\centering
\includegraphics[width=0.8\linewidth]{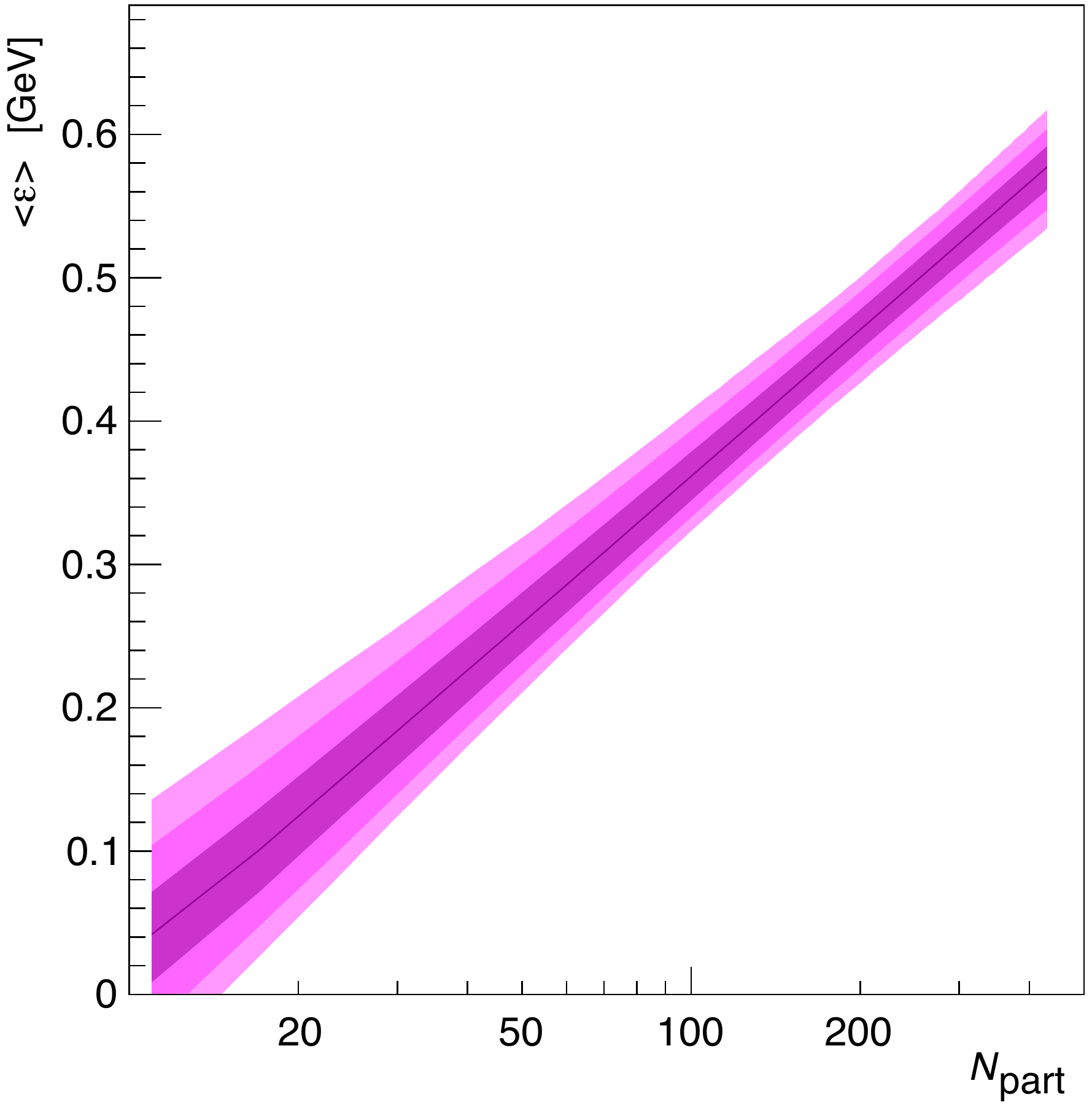}
\caption{Fitted average energy shift $\langle \epsilon \rangle$ as a function of $N_\mathrm{part}$.}
\label{fig:results_vs_Npart}
\end{figure}
Figure~\ref{fig:results_vs_Npart} shows that the fitted $\langle \epsilon \rangle$ parameter
grows logarithmically with $N_\mathrm{part}$, reaching $566\pm 15$\,MeV for $N_\mathrm{part}=400$.
The fitted value of $\sigma_\epsilon$ is $30\pm 5$\,MeV. 

\begin{figure*}[!t]
\begin{center}
\includegraphics[width=0.76\linewidth]{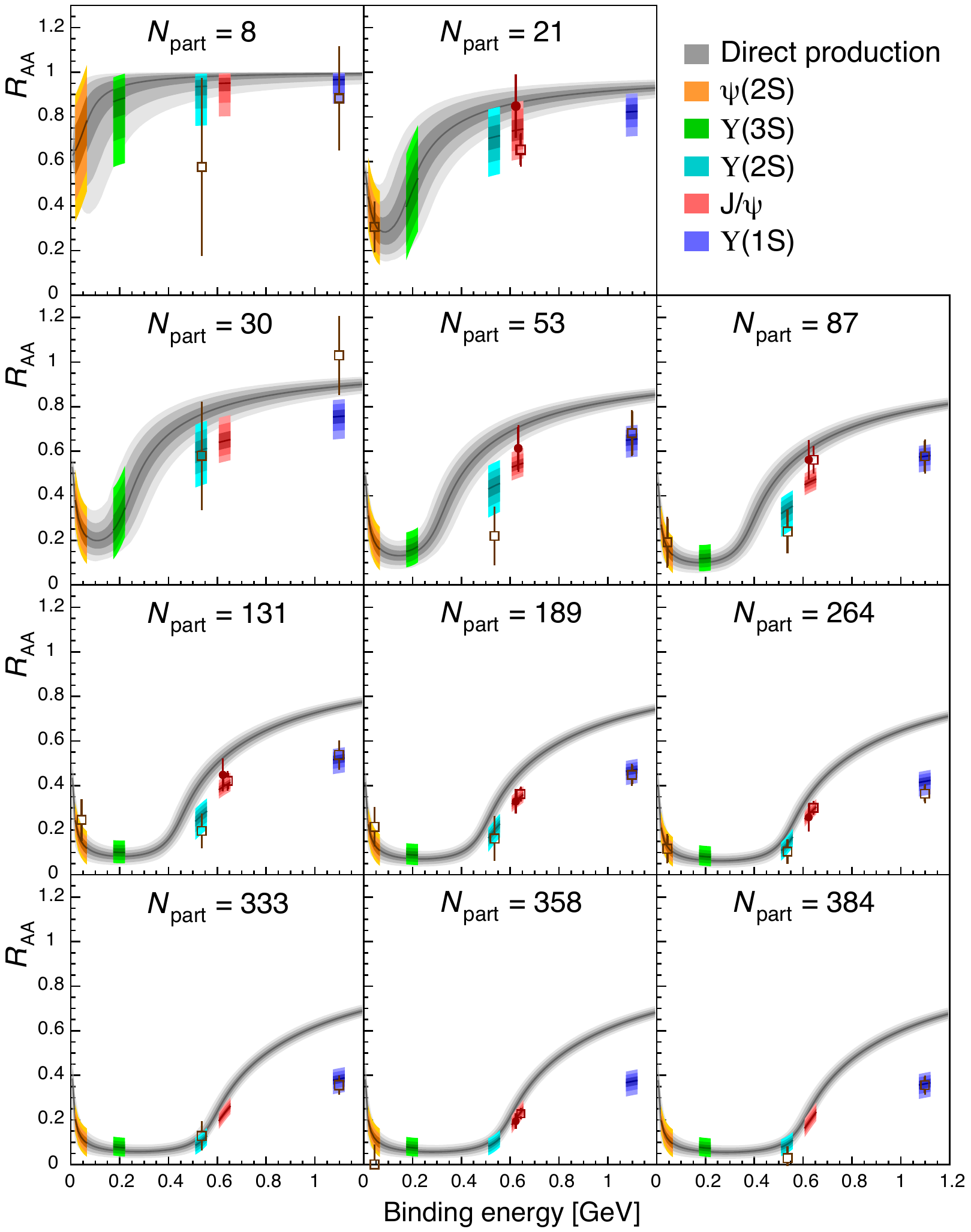}
\caption{Comparison, as a function of the binding energy, between
the data reported by CMS (closed markers) and ATLAS (open markers), 
at approximately the same average $N_\mathrm{part}$, 
and the corresponding global-fit results 
for $R_{AA}^{\mathrm{inc}}$ (coloured band segments) 
and $R_{AA}^{\mathrm{dir}}$ (continuous grey bands).}
\label{fig:results_vs_Ebinding}
\end{center}
\end{figure*}

The results of the analysis for the suppression factors are presented in Fig.~\ref{fig:results_vs_Ebinding}, 
both at the level of the direct production ($R_{AA}^{\mathrm{dir}}$), as continuous grey bands, 
and after including the effect of the feed-down contributions ($R_{AA}^{\mathrm{inc}}$),
as discreet coloured bands, one for each of the five S-wave states.
The results are shown as a function of the binding energy and in 11 bins of $N_\mathrm{part}$.
The sequence of panels shows the evolution, with increasing $N_\mathrm{part}$,
of the $E_\mathrm{b}$ dependence of $R_{AA}^{\mathrm{dir}}$, 
with its characteristic strong-suppression plateau. 
The data points are compared to the corresponding $R_{AA}^{\mathrm{inc}}$ predictions, 
illustrating how the model, with its state-independent formulation, 
is able to simultaneously reproduce the measurements reported for the several different states, 
throughout the probed spectrum of collision centrality.

The analysis of the suppression patterns measured by CMS for 
the four 1S and 2S states in 2.76\,TeV Pb-Pb collisions~\cite{bib:CMS_RAA_psi_276,bib:CMS_RAA_upsilon_276}
leads to completely analogous results, 
with slightly lower (and less precisely determined) $\langle \epsilon \rangle$ values.

\section{Discussion}

The question addressed in this paper is the following: 
is there a clear sequential pattern in the correlation between the measured nuclear suppressions 
and the binding energies of the quarkonium states?

At first sight, there is no evidence of a sequential hierarchy of suppressions: 
the data indicate that the \upsTwo is as much suppressed as the \psip, 
at least in the most central collisions, despite its $\sim$\,12 times higher binding energy.
Another relevant observation is that the relative suppression of the \psip with respect to 
the \jpsi is seemingly independent of the collision centrality in the range covered by the 
existing measurements ($N_\mathrm{part} > 20$).
If the nuclear suppression factors would be defined with respect to the 
most peripheral of the (available) Pb-Pb data ($R_{CP}$), instead of the pp data ($R_{AA}$), 
one might wrongly infer that the \psip and \jpsi are equally suppressed.

Our model of how quarkonium production is modified in nucleus-nucleus collisions 
assumes that, indeed, the nuclear suppression depends sequentially on the 
binding energy of the quarkonium states.
The starting point is the significant correlation patterns 
observed~\cite{bib:EPJCscaling} in the precise and detailed 7 and 13\,TeV pp 
data~\cite{bib:ATLASYnS,bib:ATLASpsi2S,bib:ATLASchic,bib:CMSjpsi,bib:CMSYnS,bib:BPH15005}: 
S-wave quarkonium production in pp collisions can be parametrized 
assuming that the transition probability from the pre-resonant \QQbar state 
to the physically observable bound state 
is simply proportional to a power law in the binding energy, 
$P(\QQbar \to \mathcal{Q})_\mathrm{pp} \propto E_\mathrm{b}^\delta$, equal for all states. 
We use this parametrization, extended to the P-wave states, 
as a reference description of pp quarkonium production, 
including the detailed structure of the indirect production via feed-down decays. 
The nuclear suppression effect is then modelled by a minimal modification of the pp reference formula,
introducing a threshold mechanism parametrizable with a ``penalty'' applied to the binding energy: 
$P(\QQbar \to \mathcal{Q})_{AA} \propto (E_\mathrm{b}-\epsilon)^\delta$, 
with $P = 0$ for $E_\mathrm{b} < \epsilon$, 
where $\epsilon$ depends on collision energy and centrality 
but is \emph{identical for all $c\bar{c}$ and $b\bar{b}$ states}.

A global fit to ATLAS and CMS data at 5.02\,TeV shows that this hypothesis  
describes quantitatively the observed suppression patterns for the  different final states, 
accounting for the similarity of the \upsTwo and \psip suppressions in central collisions 
and for the apparent constancy of the relative suppression of the \psip with respect to the \jpsi. 
This latter effect is actually explained by the fact that the average binding-energy penalty 
$\langle \epsilon \rangle$, a ``thermometer'' of the global nuclear modification effect, 
increases \emph{logarithmically} with the number of participants $N_\mathrm{part}$, 
while the data points, distributed uniformly in $N_\mathrm{part}$, 
mostly populate an asymptotic maximum-suppression region. 
By adopting a binning in $\ln(N_\mathrm{part})$, 
future measurements may provide a more complete view of the suppression effect, 
being the domain of the most peripheral collisions 
the one where the most characterizing $N_\mathrm{part}$ dependence should be observed.

\section{Summary}

Despite the lack of experimental information on both the production (in pp collisions) 
and the suppression (in Pb-Pb collisions) of all the P-wave states, 
with their crucial and mostly unknown feed-down contributions 
to the (inclusive) production of the S-wave states, 
the available measurements already provide a very good starting point 
for a detailed investigation of how the nuclear effects depend on the \QQbar binding energy,
over the maximally wide physical range $E_\mathrm{b} \simeq 50$--1100\,MeV.

The conclusion of our study is that the measurements provide evidence of a sequential nuclear suppression, 
which increasingly penalizes the production of the more weakly bound states, 
as foreseen~\cite{bib:DPS, bib:KKS}
when the mechanism at play is a screening of the binding forces inside the quark gluon plasma.
In first approximation, no additional nuclear effects are needed to describe 
the measured suppression patterns, integrated in the \pt and $|y|$ domains probed by the
CMS and ATLAS data. 
Such effects might become visible in more detailed multi-dimensional analyses, 
also including the dependences of the suppression rates on \pt and $y$, 
to be performed once the experimental data will become more precise and/or will 
cover an extended phase space region. 
For instance, including accurate measurements of all S-wave quarkonia
in the forward rapidity region covered by ALICE and LHCb
should allow us to disentangle the impact of the nuclear effects on the parton distribution functions.

It is particularly remarkable that a single binding-energy scaling pattern 
seamlessly reproduces the state dependence of the quarkonium yield in pp collisions 
and also, with the additional ``penalty'' effect induced by the medium, in nucleus-nucleus collisions. 
This result consolidates the success and relevance of the simple, 
factorizable and universal description of quarkonium production 
discussed in Refs.~\cite{bib:EPJCscaling,bib:simplePatterns}, 
further urging investigations on its consistency with non-relativistic QCD~\cite{bib:NRQCD}
and on the fundamental origin of such unexpected simplicity,
as recently argued in Ref.~\cite{bib:chic}.

\bigskip
Acknowledgement: 
The proton-proton baseline model that we use as a starting point for the study described in this paper
was developed in collaboration with M.\ Ara\'ujo and J.\ Seixas.


\end{document}